\documentclass[aip,twocolumn,10pt, floatfix,final]{revtex4-1}

\usepackage[caption=false]{subfig}
\usepackage{graphicx, amsmath, bm, amsfonts, amssymb, color}
\usepackage[%
  pdftex,
  breaklinks=true,
  colorlinks=true,
  linkcolor=blue,
  urlcolor=blue,
  citecolor=blue
]{hyperref}
\usepackage{verbatim}
\usepackage{xcolor}
\graphicspath{ {./images/} }

\newcommand{\beq}[0]{\begin{equation}}
\newcommand{\eeq}[0]{\end{equation}}

\usepackage{enumitem}

\newcommand{\Ch}[1]{\color{black}#1\color{black}}

\begin{document}

\title{Model for 1/f Flux noise in Superconducting Aluminum Devices: Impact of External Magnetic Fields}
\author{Jos\'{e} Alberto Nava Aquino}
\thanks{Authors to whom correspondence should be addressed: \href{mailto:rdesousa@uvic.ca}{rdesousa@uvic.ca} and \href{mailto:albertonava@uvic.ca}{albertonava@uvic.ca}}
\author{Rog\'{e}rio de Sousa}
\thanks{Authors to whom correspondence should be addressed: \href{mailto:rdesousa@uvic.ca}{rdesousa@uvic.ca} and \href{mailto:albertonava@uvic.ca}{albertonava@uvic.ca}}
\affiliation{Department of Physics and Astronomy, University of Victoria, Victoria, British Columbia V8W 2Y2, Canada}
\affiliation{Centre for Advanced Materials and Related Technology, University of Victoria, Victoria, British Columbia V8W 2Y2, Canada}

\date{\today}

\begin{abstract}
Superconducting quantum interference devices (SQUIDs) and related circuits made of aluminum are known to display $1/\omega$ flux noise, where $\omega$ is frequency. A recent experiment showed that the application of an external magnetic field in the $10-100$~G range changed the noise to a single Lorentzian peaked at $\omega=0$. Here it is shown that a model based on independent impurity spin flips with coexisting cross and direct mechanisms of spin relaxation may explain these experiments. The model shows that application of an external magnetic field can be used to reduce the impact of flux noise in qubits. 
\end{abstract}
\maketitle

Superconducting qubits are a promising candidate for building practical quantum computers due to their relatively low fabrication costs, low decoherence and scalability compared to other qubit technologies.\cite{Kjaergaard2019, you2015}
Frequency tunability is required to scale up to thousands of qubits and prevent errors associated to frequency crowding. This requires the addition of SQUIDs to the circuits, making them sensitive to flux noise. \cite{Hutchings2017} Flux noise can alter the energy levels of the qubits and lead to errors in quantum operations, making it a crucial factor in the development of practical quantum computers. \cite{Oliver2013, Wellstood1987}

Despite extensive research, the underlying mechanism of flux noise remains unclear, making it difficult to mitigate its impact. However, flux noise is widely believed to arise from the dynamics of magnetic impurities near the superconducting wires. \cite{deSousa2007,Koch2007a, Sendelbach2008, Faoro2008, Kumar2016, deGraaf2017, Quintana2017}

Several experiments \cite{Bylander2011, Anton2013, Lanting2014a, Quintana2017, Zaborniak2021} show that the frequency dependence of flux noise follows a power law, $\tilde{S}_{\Phi}(\omega)\propto 1/\omega^{\alpha}$, with exponent $\alpha$ independent of frequency $\omega$ over several decades. Experimental measurements on Nb devices show $\alpha(T)$ going from $0.8$ to $0.4$ with increasing temperature $T$. \cite{Anton2013} On the other hand, in Al devices, $\alpha=0.96-1.05$.\cite{Quintana2017} Recently, we showed that the $\alpha(T)$ measured in Nb devices can be explained by a model that assumes the impurity spins interact via ferromagnetic exchange interactions. \cite{Nava2022} In contrast, the exponent $\alpha$ measured in Al devices could not be explained by spin-spin exchange; it required instead the assumption of an extra individual spin relaxation channel, presumably due to the interaction between each spin and other non-spin degrees of freedom such as phonons, electron gas, or amorphous two-level systems. \cite{Nava2022} 

A recent experiment, focused in Al superconducting qubits,  presented for the first time measurements of flux noise in the presence of an external magnetic field $B$. \cite{Rower2023} 
While at $B=0$ the flux noise showed the expected $1/\omega$ frequency dependence \cite{Quintana2017} consistent with our model for spin relaxation disorder, \cite{Nava2022} the measurements at weak fields ($B=10-100$~G) were quite puzzling. At $B>0$ the noise transitioned smoothly into a Lorentzian in frequency, suggesting dramatic reduction in spin disorder.


Here we propose a model for these observations, and show that fitting to experimental data will shed light on the microscopic mechanism responsible for flux noise in Al devices.



Consider a set of spin impurities spatially distributed on the surfaces and interfaces of the wires forming the superconducting device. 
Each impurity is located at position $\bm{R}_j$ and has spin described by the dimensionless operator $\hat{\bm{s}}_j$. Their magnetic moment $-g\mu_B \hat{\bm{s}}_j$ imprints a flux on the device \cite{LaForest2015}
\beq
\hat{\Phi}=-\sum_j\bm{F}_j\cdot \hat{\bm{s}}_j,
\label{Phi}
\eeq
where the flux vector $\bm{F}_j=g\mu_B\bm{B}_{I}(\bm{R}_j)/I$ accounts for the dependence of the impurity-generated flux on different spin orientations. Here $g$ is the impurity's $g$-factor, $\mu_B$ is the Bohr magneton, $\bm{B}_{I}(\bm{R}_j)$ is the magnetic field generated by the wire at the spin site, and $I$ is the total current flowing through the wire. 

The spins themselves are coupled to the wire's current according to the Hamiltonian,
\beq
{\cal H} = g\mu_B  \sum_j \bm{B}_j \cdot \hat{\bm{s}}_j,
\label{spin_hamiltonian}
\eeq
where 
$\bm{B}_j=\bm{B}_I(\bm{R}_j)+\bm{B}_{{\rm ext}}$ is the spin's local field, including an externally applied $\bm{B}_{{\rm ext}}$.  For typical SQUIDs, $I\lesssim 1$~$\mu$A, leading to peak $B_I<\mu_0I/b\sim 0.1$~G where $b\sim 0.1$~$\mu$m is the thickness of the superconducting wire. As a result it is safe to approximate $\bm{B}_j\approx\bm{B}_{{\rm ext}}$ when $B_{{\rm ext}}>1$~G. 

Flux noise arises from time-dependent correlations of the flux fluctuation operator $\delta\hat{\Phi}(t)=\hat{\Phi}(t)-\langle \hat{\Phi}\rangle$,
\begin{eqnarray} \label{eq_fluxnoise}
\tilde{S}_{\Phi} (\omega) &=& \int_{-\infty}^{\infty}dt \textrm{e}^{i\omega t} \left\langle \delta \hat{\Phi}(t)\delta\hat{\Phi}(0)\right\rangle\nonumber\\
&=&\sum_{j,k,a,b} F^{a}_j \tilde{S}^{ab}_{jk} (\omega) F^{b}_k,
\end{eqnarray}
where the superscripts $a,b=x,y,z$ denote the components of the flux vector, and the spin noise is defined as
\begin{equation}
\tilde{S}^{ab}_{jk} (\omega) = \int_{-\infty}^{\infty} dt e^{i \omega t} \langle [\hat{s}_j^{a}(t) - \langle \hat{s}_j^{a} \rangle] [\hat{s}_k^{b}(0) - \langle \hat{s}_k^{b} \rangle]\rangle.
\label{spinnoise}
\end{equation}

In Ref.~\onlinecite{Nava2022} a numerical method to compute Eq.~(\ref{spinnoise}) for a general Hamiltonian of interacting spins was described. 
While it was shown that spin-spin interactions lead to the dominant mechanism of flux noise in Nb devices, the so called spin diffusion mechanism, comparison with experiments in Al devices suggested a quite different picture. 
Namely, that the spins near Al wires were fluctuating independently with spin-spin interaction and spin diffusion playing a minor role. 


In the current letter, it is assumed that independent spin flips dominate $\omega>0$ noise, so that $\tilde{S}_{jk}(\omega)\ll \tilde{S}_{jj}(\omega)$ when $j\neq k$.
Together with $\tilde{S}^{ab}_{jj}(\omega)=-\tilde{S}^{ba}_{jj}(\omega)$ when $a\neq b$, \cite{NoteSymmetry} the flux noise simplifies to
\beq
\tilde{S}_{\Phi}(\omega)\approx\sum_j \left[\left|\bm{F}_j\cdot \hat{\bm{B}}_j\right|^{2}\tilde{S}^{\parallel}_{jj}(\omega)+\left|\bm{F}_j\times \hat{\bm{B}}_j\right|^{2}\tilde{S}^{\perp}_{jj}(\omega)\right].
\label{SPhiIndependenSpin}
\eeq
Here $\hat{\bm{B}}_j$ is the unit vector pointing along the local magnetic field acting on the spin, $\tilde{S}^{\parallel}_{jj}(\omega)\equiv \tilde{S}^{\hat{B}_j\hat{B}_j}_{jj}(\omega)$, and  $\tilde{S}^{\perp}_{jj}(\omega)\equiv \tilde{S}^{\hat{P}_j\hat{P}_j}_{jj}(\omega)$ were $\hat{\bm{P}}_j$ is any direction perpendicular to $\bm{B}_j$. 

\begin{figure}
\begin{center}
\includegraphics[width=.99\linewidth]{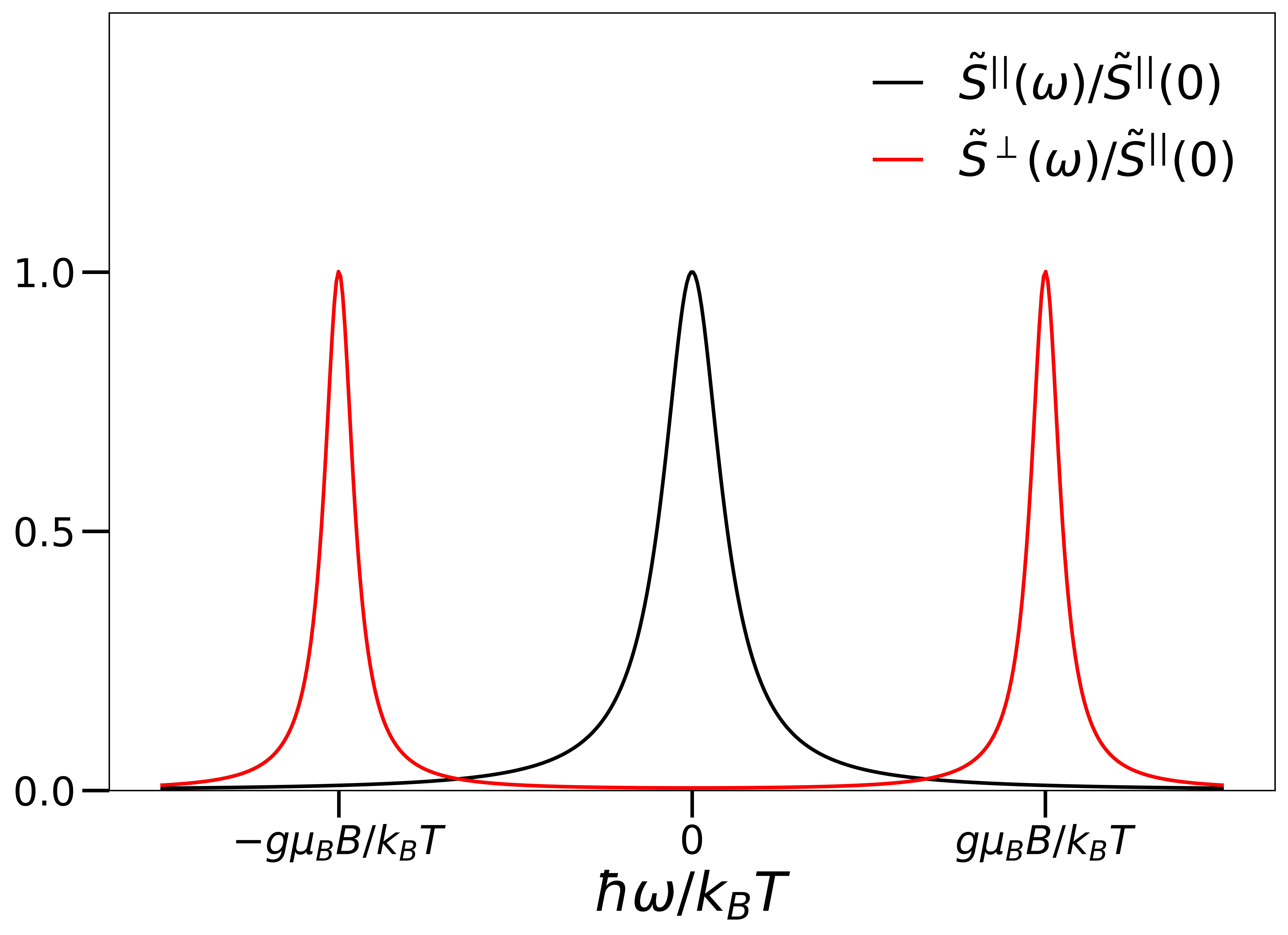}
\end{center}
\caption{Flux noise produced by a single spin in a magnetic field $\bm{B}$. Spin fluctuations along $\bm{B}$ ($\tilde{S}^{\parallel}(\omega)$) remain peaked at $\omega=0$. In contrast, spin fluctuations perpendicular to $\bm{B}$ ($\tilde{S}^{\perp}(\omega)$) give rise to ``spin precession peaks'' centered at $\omega=\pm g\mu_BB/\hbar$. As a result the flux noise contribution due to the component of the flux vector perpendicular to $\bm{B}$ is expelled from the low frequency range (See Eq.~(\ref{SPhiIndependenSpin})). 
Both contributions are Lorentzian with linewidth set by the corresponding spin decay rate $\Gamma_{\parallel}, \Gamma_{\perp}$. \Ch{The figure assumes $\Gamma_{\parallel}=2\Gamma_{\perp}=0.1\;g\mu_B B/\hbar$}.
\label{fig:fig1}}
\end{figure}

\begin{figure}
\begin{center}
\includegraphics[width=.99\linewidth]{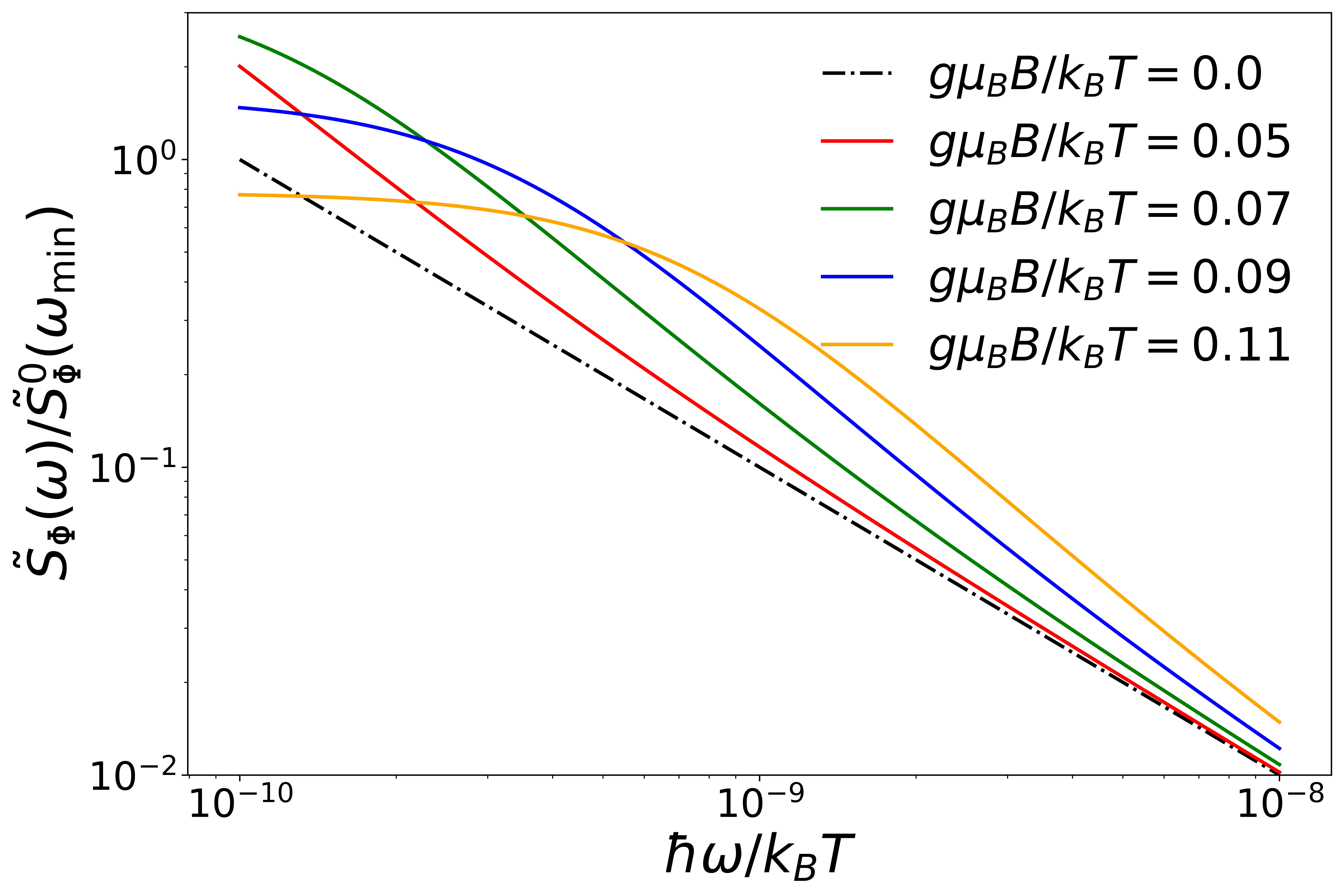}
\end{center}
\caption{Effect of external magnetic field $\bm{B}$ on flux noise with $\bm{F}_j=\bm{F}\parallel \bm{B}$. The spin relaxation rates are assumed to be distributed according to Eq.~(\ref{Gamma_model1}), with exponent $n=4$ representing direct phonon emission due to spin-orbit admixture. Flux noise transitions from $1/\omega$ ``disorder'' noise to an ``undisordered'' Lorentzian with increasing $B$. The noise is normalized by $\tilde{S}_{\Phi}^0 (\omega_{{\rm min}})$, its value at $B=0$ and $\hbar \omega_{{\rm min}}/k_BT = 10^{-10}$. Parameters used in the calculation: $\hbar \Gamma_0/ k_B T = 1, \hbar \Gamma_1= 5 \times10^{-6} (k_B T)^{-4}, \lambda_{{\rm max}} = 30$. 
\label{fig:fig2}}
\end{figure}


It is assumed that each spin evolves independently of the others according to the equation of motion
\begin{eqnarray} \label{eom_non-int}
    \frac{d \bm{s}}{d t} &=&  \frac{g\mu_B}{\hbar} \bm{B} \times \bm{s}  
    \Ch{-\Gamma_{\parallel}[(\bm{s} - \bm{s}^{{\rm inst~eq}})\cdot \hat{\bm{B}}] \hat{\bm{B}}}\nonumber\\
    &&\Ch{-\Gamma_{\perp}[\hat{\bm{B}}\times(\bm{s} - \bm{s}^{{\rm inst~eq}})] \times \hat{\bm{B}}},
\end{eqnarray}
where $\bm{s}=\langle \hat{\bm{s}}_j\rangle$ represents a quantum average (the subscript $j$ is now omitted for simplicity).
In addition to spin precession, this equation includes  \Ch{\emph{spin energy relaxation rate} 
$\Gamma_{\parallel}=\frac{1}{T_1}$ and \emph{spin decoherence rate} $\Gamma_{\perp}=\frac{1}{T_2}\ge \Gamma_{\parallel}/2$. These drive $\bm{s}$ towards its ``instantaneous equilibrium'' value $\bm{s}^{{\rm inst~eq}}=\bm{s}^{{\rm eq}}-g\mu_B[\tilde{\chi}_{\parallel}(0)\delta\bm{B}_{\parallel}(t)+\tilde{\chi}_{\perp}(0)\delta\bm{B}_{\perp}(t)]$, where $\tilde{\chi}_{\parallel}(0)$ and $\tilde{\chi}_{\perp}(0)$ are the $\omega=0$ longitudinal and transverse susceptibilities, respectively}. For small deviations from equilibrium we write $\bm{B}\rightarrow \bm{B}+\delta\bm{B}(t)$, and $\bm{s} \rightarrow \bm{s}^{{\rm eq}} + \delta \bm{s}(t)$, where both $\delta\bm{B}(t)$ and $\delta\bm{s}(t)$ are small time-dependent perturbations. Plugging these into Eq.~(\ref{eom_non-int}) and dropping non-linear terms enables the calculation of the dynamical susceptibility and the spin noise through the fluctuation-dissipation theorem (See~\onlinecite{Nava2022} for details). In the current independent spin model the spin noise can be calculated analytically:
\begin{subequations}
\begin{eqnarray}
    \tilde{S}^{\parallel}(\omega) &=& \frac{2 \hbar \omega}{1 - e^{-\hbar \omega /k_B T}} \frac{\Gamma_{\parallel} \tilde{\chi}_{\parallel}(0)}{\omega^2 + \Gamma_{\parallel}^{2}},\label{sn_zz_non-int}\\
    \Ch{\tilde{S}^{\perp}(\omega)} &\Ch{=}& \Ch{\frac{\hbar\omega\Gamma_{\perp}\tilde{\chi}_{\perp}(0)}{1 - e^{-\hbar \omega /k_B T}}
    \left[ \frac{1}{(\omega - g\mu_B B/\hbar)^2 + \Gamma_{\perp}^{2}}\right.}\nonumber\\
    &&\Ch{+ \left.\frac{1}{(\omega + g\mu_B B/\hbar)^2 + \Gamma_{\perp}^{2}} \right]}\label{sn_xx_non-int},
%
\end{eqnarray}
\end{subequations}
where $s^{{\rm eq}} =-\frac{1}{2} {\rm tanh}(g\mu_B B/2 k_B T)$ and \Ch{$\tilde{\chi}_{\parallel}(0) = 1/[4k_B T\cosh^{2}(g\mu_BB/2k_BT)]$, $\tilde{\chi}_{\perp}(0) = {\rm tanh}(g\mu_B B/2 k_B T)/(2g\mu_BB)$}.

In the presence of an external magnetic field, $\tilde{S}^{\parallel}(\omega)$ is drastically different from $\tilde{S}^{\perp}(\omega)$, as illustrated in Fig.~\ref{fig:fig1}. From Eq.~(\ref{SPhiIndependenSpin}) it follows that at each spin location, the flux vector's component parallel to the external field will produce Lorentzian flux noise peaked at $\omega=0$. 
In contrast, the component of the flux vector perpendicular to $\bm{B}$ will produce instead two Lorentzians peaked at $\omega= \pm g\mu_B B/\hbar=\pm 2\pi (B/1~{\rm G})\times 2.8~{\rm MHz}$. This shows that even fields of a few Gauss have huge impact on flux noise: They shift the perpendicular flux vector contribution to a much higher frequency range. 

\Ch{This effect is the spin analogue of high-frequency charge noise produced by quantum coherent two-level systems in amorphous materials.\cite{Shnirman2005}}


While a single spin produces Lorentzian noise, a system of spins with a wide distribution of relaxation rates $\Gamma_{\parallel}$ can give rise to $1/\omega$ noise in a large interval of frequencies. \cite{deSousa2007, Belli2020, Nava2022}

Motivated by recent flux noise measurements in Al devices under weak magnetic fields, \cite{Rower2023} we propose the following relaxation rates for a spin $\bm{s}_j$ placed on the surface or interface of a superconducting wire,
\beq \label{Gamma_model1}
    \Gamma_{\parallel, j}= \Gamma_0 e^{-\lambda_j} +    
    \gamma (k_BT) (g\mu_B B)^n.
\eeq
The first contribution to Eq. (\ref{Gamma_model1}) models \emph{cross spin relaxation}  due to the impurity spin interaction with one or more amorphous two-level systems (TLSs), where $\lambda_j$ is a random variable uniformly distributed in the interval $[0,\lambda_{{\rm max}}]$. \cite{deSousa2007, Belli2020} Cross relaxation occurs when a TLS switches due to thermal fluctuation, causing a simultaneous impurity spin flip. Since the majority of these processes occur for TLSs with energy splitting of the order of $k_BT$, cross relaxation is independent of magnetic field when $g\mu_B B< k_BT$. \cite{deSousa2007} 
The second contribution to Eq.~(\ref{Gamma_model1}) models \emph{direct spin relaxation}, whereby the spin emits a phonon, either due to modulation of its hyperfine interaction ($\Gamma_{\parallel,j}\propto T\;B^2$), or due to the spin-orbit admixture mechanism ($\Gamma_{\parallel, j}\propto T\;B^4$). In both cases the rates scale linearly in $T$ because they are proportional to the phonon occupation at the Zeeman frequency; they also scale as $B^n$ because the direct mechanism requires the breaking of time-reversal symmetry. \cite{VanVleck1940, deSousa2003b} The constants $\Gamma_0$, \Ch{$\gamma$}, and $n$ are to be determined by fitting to experiments.


The $\Gamma_{\parallel, j}$ are distributed according to probability density 
$p(\Gamma) = 1/(\lambda_{{\rm max}} |\frac{d \Gamma}{d \lambda}|)$, leading to the disorder-averaged single spin noise, 
%
%
%
%
\begin{widetext}
\begin{eqnarray}
\tilde{S}_{{\rm avg}}^{\parallel}(\omega) &=& \int_{\Gamma_{{\rm min}}}^{\Gamma_{{\rm max}}} \tilde{S}^{\parallel}(\omega) p(\Gamma) d \Gamma  =  \Ch{\frac{ \hbar \omega \tilde{\chi}_{\parallel}(0)}{1 - e^{-\hbar \omega /k_B T}} 
\left\{ 
\frac{1}{\lambda_{{\rm max}}}\left(\frac{\omega^2}{\omega^2+\Gamma_{B}^{2}}\right)
\frac{{\rm arctan}(\Gamma_{{\rm max}}/\omega)-{\rm arctan}(\Gamma_{{\rm min}}/\omega)}{\omega}\right.\nonumber\\}
&&\Ch{\left.+\left[1 - \frac{1}{2\lambda_{{\rm max}}} {{\rm log}} \left( \frac{\Gamma_{{\rm max}}^2 + \omega^2}{\Gamma_{{\rm min}}^2 + \omega^2 }\right)\right]
\frac{\Gamma_B}{\omega^{2}+\Gamma_{B}^{2}}\right\}},
\label{Szz_avg}
\end{eqnarray}
\end{widetext}
where \Ch{$\Gamma_B = \gamma (k_BT)(g\mu_BB)^n$}~and
the minimum and maximum rates are 
$\Gamma_{{\rm min}}=\Gamma_0 e^{-\lambda_{{\rm max}}} + \Gamma_B$ and $\Gamma_{{\rm max}} = \Gamma_0 + \Gamma_B$.



When $B=0$, \Ch{$\Gamma_B=0$ and the first term in}~the curly bracket of Eq. (\ref{Szz_avg}) gives rise to $1/|\omega|$ noise for a wide range of frequencies $\Gamma_{{\rm min}} \ll |\omega| \ll \Gamma_{{\rm max}}$. As $B$ is increased from zero, 
%
%
%
there will be a range of frequencies satisfying $|\omega| < \Gamma_B\approx \Gamma_{{\rm min}}$. \Ch{In this range the $1/|\omega|$ contribution to flux noise is suppressed, and the second term in the curly bracket of 
Eq.~(\ref{Szz_avg}) becomes important, with the square bracket approximately independent of $\omega$.
As $B$ increases further $\Gamma_{{\rm min}}$ gets close to $\Gamma_{{\rm max}}$ and the low frequency noise is converted into a simple ``reduced disorder'' Lorentzian, centered at $\omega=0$ with width $\Gamma_B$}.

%
%
%
%
%

Figure~\ref{fig:fig2} shows the predicted transition from $1/\omega$ to Lorentzian flux noise  with increasing magnetic field. The behavior is in qualitative agreement to recent experiments in Al devices (Fig. 3(a) of Ref.~\onlinecite{Rower2023}). \Ch{Our Eqs.~(\ref{SPhiIndependenSpin})~and~(\ref{Szz_avg}) can explain the experimental data if we assume 
$\lambda_{{\rm max}} =10-30$ and spin areal density $3-8 \times 10^{16}/{\rm m}^2$}. 

\Ch{The spin precession peaks at $\omega =\pm g\mu_B B/\hbar$, the ones that occur due to the nonzero components of $\bm{F}_j\perp \bm{B}$,
may also broaden due to disorder. If these are dominated by spin energy relaxation, we get $\Gamma_{\perp,j}\approx \frac{1}{2}\Gamma_{\parallel,j}$ and the peaks transition from $\propto 1/|\omega\pm g\mu_B B/\hbar|$ at low $B$ to Lorentzians centered at $\pm g\mu_B B/\hbar$ with width $\Gamma_B$ at higher $B$. Measuring the shape of these spin precession peaks will yield information about the distribution of spin decoherence rates $\Gamma_{\perp,j}$}.


In conclusion, a model for spin noise is proposed to explain flux noise experiments in superconducting Al devices. The model is based on the assumption that the interaction between each impurity spin and the vibrational modes of the lattice, formed by amorphous TLSs and phonons, dominate finite frequency fluctuations. Within this scenario, the impact of spin-spin interactions such as dipolar and exchange can still be observed as a Curie-Weiss temperature $T_{{\rm CW}}$ in the zero-frequency susceptibility, because decay rates $\Gamma_{\parallel}, \Gamma_{\perp}$ do not impact $\tilde{\chi}_{\parallel}(0)$. Thus one can use $\tilde{\chi}_{\parallel}(0)= 1/[4k_B(T-T_{{\rm CW}})\cosh^{2}(g\mu_BB/2k_BT)]$ in Eq.~(\ref{Szz_avg}) to model spin-spin interactions.

An analytical expression for the flux noise in the presence of an external magnetic field $B$ is obtained, revealing a transition from $1/\omega$ flux noise at $B=0$ to Lorentzian flux noise in the frequency range where direct spin relaxation by phonon emission is stronger than cross relaxation due to amorphous TLSs. The transition is similar to what is observed in recent experiments.\cite{Rower2023}

Fitting the current theory to experimental data will yield the characteristic exponent $n$ for direct phonon emission, elucidating whether the mechanism is mediated by spin-orbit coupling or hyperfine interaction. This will in turn yield valuable information on the identity of the spins causing flux noise. 

The model predicts that application of a $B$ field reduces low frequency noise arising from $\bm{F}_j\parallel \bm{B}$ \Ch{($\bm{F}_j\perp \bm{B}$) } by a factor of $1/\cosh^{2}{(g\mu_B B/2k_BT)}$ \Ch{(${\rm tanh}(g\mu_B B/2 k_B T)/(2g\mu_BB)$)}. It also shifts the contribution of $\bm{F}_j\perp \bm{B}$ out of the low frequency range, transferring noise to spin precession peaks centered at $\omega=\pm g\mu_B B/\hbar$. \Ch{For $T=10$~mK and 
$B=100$~G, $1/\cosh^{2}{(g\mu_B B/2k_BT)} \approx 2/3$, and approximately one half of the total noise power will be transferred to a higher frequency band. Therefore, a total reduction of  $\approx 1/3$ is made to the noise power in the low frequency band. Further reduction can be achieved for $B> 100$~G. These results show that } an external magnetic field can be used to reduce the impact of flux noise in qubits.




\begin{acknowledgments}
The authors thank M. Amin, R. Harris, and T. Lanting for several insights into flux noise, and L. Ateshian and D. A. Rower for discussions and sharing experimental results before publication. This work was supported by NSERC (Canada) through its Discovery program (Grant number RGPIN-2020-04328).
\end{acknowledgments}

\bibliography{fluxnoise_Alberto}

\begin{thebibliography}{24}%
\makeatletter
\providecommand \@ifxundefined [1]{%
 \@ifx{#1\undefined}
}%
\providecommand \@ifnum [1]{%
 \ifnum #1\expandafter \@firstoftwo
 \else \expandafter \@secondoftwo
 \fi
}%
\providecommand \@ifx [1]{%
 \ifx #1\expandafter \@firstoftwo
 \else \expandafter \@secondoftwo
 \fi
}%
\providecommand \natexlab [1]{#1}%
\providecommand \enquote  [1]{``#1''}%
\providecommand \bibnamefont  [1]{#1}%
\providecommand \bibfnamefont [1]{#1}%
\providecommand \citenamefont [1]{#1}%
\providecommand \href@noop [0]{\@secondoftwo}%
\providecommand \href [0]{\begingroup \@sanitize@url \@href}%
\providecommand \@href[1]{\@@startlink{#1}\@@href}%
\providecommand \@@href[1]{\endgroup#1\@@endlink}%
\providecommand \@sanitize@url [0]{\catcode `\\12\catcode `\$12\catcode
  `\&12\catcode `\#12\catcode `\^12\catcode `\_12\catcode `\%12\relax}%
\providecommand \@@startlink[1]{}%
\providecommand \@@endlink[0]{}%
\providecommand \url  [0]{\begingroup\@sanitize@url \@url }%
\providecommand \@url [1]{\endgroup\@href {#1}{\urlprefix }}%
\providecommand \urlprefix  [0]{URL }%
\providecommand \Eprint [0]{\href }%
\providecommand \doibase [0]{http://dx.doi.org/}%
\providecommand \selectlanguage [0]{\@gobble}%
\providecommand \bibinfo  [0]{\@secondoftwo}%
\providecommand \bibfield  [0]{\@secondoftwo}%
\providecommand \translation [1]{[#1]}%
\providecommand \BibitemOpen [0]{}%
\providecommand \bibitemStop [0]{}%
\providecommand \bibitemNoStop [0]{.\EOS\space}%
\providecommand \EOS [0]{\spacefactor3000\relax}%
\providecommand \BibitemShut  [1]{\csname bibitem#1\endcsname}%
\let\auto@bib@innerbib\@empty
\bibitem [{\citenamefont {Kjaergaard}\ \emph {et~al.}(2019)\citenamefont
  {Kjaergaard}, \citenamefont {Schartz}, \citenamefont {Braum\"{u}ller},
  \citenamefont {Krantz}, \citenamefont {Wang}, \citenamefont {Gustavsson},\
  and\ \citenamefont {Oliver}}]{Kjaergaard2019}%
  \BibitemOpen
  \bibfield  {author} {\bibinfo {author} {\bibfnamefont {M.}~\bibnamefont
  {Kjaergaard}}, \bibinfo {author} {\bibfnamefont {M.~E.}\ \bibnamefont
  {Schartz}}, \bibinfo {author} {\bibfnamefont {J.}~\bibnamefont
  {Braum\"{u}ller}}, \bibinfo {author} {\bibfnamefont {P.}~\bibnamefont
  {Krantz}}, \bibinfo {author} {\bibfnamefont {J.~I.-J.}\ \bibnamefont {Wang}},
  \bibinfo {author} {\bibfnamefont {S.}~\bibnamefont {Gustavsson}}, \ and\
  \bibinfo {author} {\bibfnamefont {W.~D.}\ \bibnamefont {Oliver}},\ }\bibfield
   {title} {\enquote {\bibinfo {title} {{Superconducting qubits: Current state
  of play}},}\ }\href {\doibase
  https://doi.org/10.1146/annurev-conmatphys-031119-050605} {\bibfield
  {journal} {\bibinfo  {journal} {Annu. Rev. Condens. Matter Phys.}\ }\textbf
  {\bibinfo {volume} {11}},\ \bibinfo {pages} {369} (\bibinfo {year}
  {2019})}\BibitemShut {NoStop}%
\bibitem [{\citenamefont {You}\ and\ \citenamefont {Nori}(2005)}]{you2015}%
  \BibitemOpen
  \bibfield  {author} {\bibinfo {author} {\bibfnamefont {J.~Q.}\ \bibnamefont
  {You}}\ and\ \bibinfo {author} {\bibfnamefont {F.}~\bibnamefont {Nori}},\
  }\bibfield  {title} {\enquote {\bibinfo {title} {{Superconducting circuits
  and quantum information}},}\ }\href {\doibase
  https://doi.org/10.1063/1.2155757} {\bibfield  {journal} {\bibinfo  {journal}
  {Phys. Today}\ }\textbf {\bibinfo {volume} {58}},\ \bibinfo {pages} {42}
  (\bibinfo {year} {2005})}\BibitemShut {NoStop}%
\bibitem [{\citenamefont {Hutchings}\ \emph {et~al.}(2017)\citenamefont
  {Hutchings}, \citenamefont {Hertzberg}, \citenamefont {Liu}, \citenamefont
  {Bronn}, \citenamefont {Keefe}, \citenamefont {Brink}, \citenamefont {Chow},\
  and\ \citenamefont {Plourde}}]{Hutchings2017}%
  \BibitemOpen
  \bibfield  {author} {\bibinfo {author} {\bibfnamefont {M.~D.}\ \bibnamefont
  {Hutchings}}, \bibinfo {author} {\bibfnamefont {J.~B.}\ \bibnamefont
  {Hertzberg}}, \bibinfo {author} {\bibfnamefont {Y.}~\bibnamefont {Liu}},
  \bibinfo {author} {\bibfnamefont {N.~T.}\ \bibnamefont {Bronn}}, \bibinfo
  {author} {\bibfnamefont {G.~A.}\ \bibnamefont {Keefe}}, \bibinfo {author}
  {\bibfnamefont {M.}~\bibnamefont {Brink}}, \bibinfo {author} {\bibfnamefont
  {J.~M.}\ \bibnamefont {Chow}}, \ and\ \bibinfo {author} {\bibfnamefont
  {B.~L.~T.}\ \bibnamefont {Plourde}},\ }\bibfield  {title} {\enquote {\bibinfo
  {title} {{Tunable Superconducting Qubits with Flux-Independent Coherence}},}\
  }\href {http://doi.org/10.1103/PhysRevApplied.8.044003} {\bibfield  {journal}
  {\bibinfo  {journal} {Phys. Rev. Appl.}\ }\textbf {\bibinfo {volume} {8}},\
  \bibinfo {pages} {044003} (\bibinfo {year} {2017})}\BibitemShut {NoStop}%
\bibitem [{\citenamefont {Oliver}\ and\ \citenamefont
  {Welander}(2013)}]{Oliver2013}%
  \BibitemOpen
  \bibfield  {author} {\bibinfo {author} {\bibfnamefont {W.~D.}\ \bibnamefont
  {Oliver}}\ and\ \bibinfo {author} {\bibfnamefont {P.~B.}\ \bibnamefont
  {Welander}},\ }\bibfield  {title} {\enquote {\bibinfo {title} {{Materials in
  superconducting quantum bits}},}\ }\href {\doibase
  https://doi.org/10.1557/mrs.2013.229} {\bibfield  {journal} {\bibinfo
  {journal} {MRS Bull.}\ }\textbf {\bibinfo {volume} {38}},\ \bibinfo {pages}
  {816} (\bibinfo {year} {2013})}\BibitemShut {NoStop}%
\bibitem [{\citenamefont {Wellstood}\ and\ \citenamefont
  {Urbina}(1987)}]{Wellstood1987}%
  \BibitemOpen
  \bibfield  {author} {\bibinfo {author} {\bibfnamefont {F.~C.}\ \bibnamefont
  {Wellstood}}\ and\ \bibinfo {author} {\bibfnamefont {J.}~\bibnamefont
  {Urbina}, \bibfnamefont {C.~amd~Clarke}},\ }\bibfield  {title} {\enquote
  {\bibinfo {title} {{ Low-frequency noise in dc superconducting quantum
  interference devices below 1 k}},}\ }\href {\doibase
  https://doi.org/10.1063/1.98041} {\bibfield  {journal} {\bibinfo  {journal}
  {Appl. Phys. Lett.}\ }\textbf {\bibinfo {volume} {50}},\ \bibinfo {pages}
  {772} (\bibinfo {year} {1987})}\BibitemShut {NoStop}%
\bibitem [{\citenamefont {{de Sousa}}(2007)}]{deSousa2007}%
  \BibitemOpen
  \bibfield  {author} {\bibinfo {author} {\bibfnamefont {R.}~\bibnamefont {{de
  Sousa}}},\ }\bibfield  {title} {\enquote {\bibinfo {title} {{Dangling-bond
  spin relaxation and magnetic $1/f$ noise from the
  amorphous-semiconductor/oxide interface: Theory}},}\ }\href {\doibase
  10.1103/PhysRevB.76.245306} {\bibfield  {journal} {\bibinfo  {journal} {Phys.
  Rev. B}\ }\textbf {\bibinfo {volume} {76}},\ \bibinfo {pages} {245306}
  (\bibinfo {year} {2007})}\BibitemShut {NoStop}%
\bibitem [{\citenamefont {Koch}, \citenamefont {DiVincenzo},\ and\
  \citenamefont {Clarke}(2007)}]{Koch2007a}%
  \BibitemOpen
  \bibfield  {author} {\bibinfo {author} {\bibfnamefont {R.~H.}\ \bibnamefont
  {Koch}}, \bibinfo {author} {\bibfnamefont {D.~P.}\ \bibnamefont
  {DiVincenzo}}, \ and\ \bibinfo {author} {\bibfnamefont {J.}~\bibnamefont
  {Clarke}},\ }\bibfield  {title} {\enquote {\bibinfo {title} {{Model for 1/f
  Flux Noise in SQUIDs and Qubits}},}\ }\href {\doibase
  10.1103/PhysRevLett.98.267003} {\bibfield  {journal} {\bibinfo  {journal}
  {Phys. Rev. Lett.}\ }\textbf {\bibinfo {volume} {98}},\ \bibinfo {pages}
  {267003} (\bibinfo {year} {2007})}\BibitemShut {NoStop}%
\bibitem [{\citenamefont {Sendelbach}\ \emph {et~al.}(2008)\citenamefont
  {Sendelbach}, \citenamefont {Hover}, \citenamefont {Kittel}, \citenamefont
  {M{\"{u}}ck}, \citenamefont {Martinis},\ and\ \citenamefont
  {McDermott}}]{Sendelbach2008}%
  \BibitemOpen
  \bibfield  {author} {\bibinfo {author} {\bibfnamefont {S.}~\bibnamefont
  {Sendelbach}}, \bibinfo {author} {\bibfnamefont {D.}~\bibnamefont {Hover}},
  \bibinfo {author} {\bibfnamefont {A.}~\bibnamefont {Kittel}}, \bibinfo
  {author} {\bibfnamefont {M.}~\bibnamefont {M{\"{u}}ck}}, \bibinfo {author}
  {\bibfnamefont {J.~M.}\ \bibnamefont {Martinis}}, \ and\ \bibinfo {author}
  {\bibfnamefont {R.}~\bibnamefont {McDermott}},\ }\bibfield  {title} {\enquote
  {\bibinfo {title} {{Magnetism in SQUIDs at Millikelvin Temperatures}},}\
  }\href {\doibase 10.1103/PhysRevLett.100.227006} {\bibfield  {journal}
  {\bibinfo  {journal} {Phys. Rev. Lett.}\ }\textbf {\bibinfo {volume} {100}},\
  \bibinfo {pages} {227006} (\bibinfo {year} {2008})}\BibitemShut {NoStop}%
\bibitem [{\citenamefont {Faoro}\ and\ \citenamefont
  {Ioffe}(2008)}]{Faoro2008}%
  \BibitemOpen
  \bibfield  {author} {\bibinfo {author} {\bibfnamefont {L.}~\bibnamefont
  {Faoro}}\ and\ \bibinfo {author} {\bibfnamefont {L.~B.}\ \bibnamefont
  {Ioffe}},\ }\bibfield  {title} {\enquote {\bibinfo {title} {{Microscopic
  origin of low-frequency flux noise in Josephson circuits}},}\ }\href
  {\doibase 10.1103/PhysRevLett.100.227005} {\bibfield  {journal} {\bibinfo
  {journal} {Phys. Rev. Lett.}\ }\textbf {\bibinfo {volume} {100}},\ \bibinfo
  {pages} {227005} (\bibinfo {year} {2008})}\BibitemShut {NoStop}%
\bibitem [{\citenamefont {Kumar}\ \emph {et~al.}(2016)\citenamefont {Kumar},
  \citenamefont {Sendelbach}, \citenamefont {Beck}, \citenamefont {Freeland},
  \citenamefont {Wang}, \citenamefont {Wang}, \citenamefont {Yu}, \citenamefont
  {Wu}, \citenamefont {Pappas},\ and\ \citenamefont {McDermott}}]{Kumar2016}%
  \BibitemOpen
  \bibfield  {author} {\bibinfo {author} {\bibfnamefont {P.}~\bibnamefont
  {Kumar}}, \bibinfo {author} {\bibfnamefont {S.}~\bibnamefont {Sendelbach}},
  \bibinfo {author} {\bibfnamefont {M.~A.}\ \bibnamefont {Beck}}, \bibinfo
  {author} {\bibfnamefont {J.~W.}\ \bibnamefont {Freeland}}, \bibinfo {author}
  {\bibfnamefont {Z.}~\bibnamefont {Wang}}, \bibinfo {author} {\bibfnamefont
  {H.}~\bibnamefont {Wang}}, \bibinfo {author} {\bibfnamefont {C.~C.}\
  \bibnamefont {Yu}}, \bibinfo {author} {\bibfnamefont {R.~Q.}\ \bibnamefont
  {Wu}}, \bibinfo {author} {\bibfnamefont {D.~P.}\ \bibnamefont {Pappas}}, \
  and\ \bibinfo {author} {\bibfnamefont {R.}~\bibnamefont {McDermott}},\
  }\bibfield  {title} {\enquote {\bibinfo {title} {{Origin and Reduction of
  $1/f$ Magnetic Flux Noise in Superconducting Devices}},}\ }\href {\doibase
  10.1103/PhysRevApplied.6.041001} {\bibfield  {journal} {\bibinfo  {journal}
  {Phys. Rev. Appl.}\ }\textbf {\bibinfo {volume} {6}},\ \bibinfo {pages}
  {041001(R)} (\bibinfo {year} {2016})}\BibitemShut {NoStop}%
\bibitem [{\citenamefont {{de Graaf}}\ \emph {et~al.}(2017)\citenamefont {{de
  Graaf}}, \citenamefont {Adamyan}, \citenamefont {Lindstr{\"{o}}m},
  \citenamefont {Erts}, \citenamefont {Kubatkin}, \citenamefont {Tzalenchuk},\
  and\ \citenamefont {Danilov}}]{deGraaf2017}%
  \BibitemOpen
  \bibfield  {author} {\bibinfo {author} {\bibfnamefont {S.~E.}\ \bibnamefont
  {{de Graaf}}}, \bibinfo {author} {\bibfnamefont {A.~A.}\ \bibnamefont
  {Adamyan}}, \bibinfo {author} {\bibfnamefont {T.}~\bibnamefont
  {Lindstr{\"{o}}m}}, \bibinfo {author} {\bibfnamefont {D.}~\bibnamefont
  {Erts}}, \bibinfo {author} {\bibfnamefont {S.~E.}\ \bibnamefont {Kubatkin}},
  \bibinfo {author} {\bibfnamefont {A.~Y.}\ \bibnamefont {Tzalenchuk}}, \ and\
  \bibinfo {author} {\bibfnamefont {A.~V.}\ \bibnamefont {Danilov}},\
  }\bibfield  {title} {\enquote {\bibinfo {title} {{Direct Identification of
  Dilute Surface Spins on {Al$_2$O$_3$}: Origin of Flux Noise in Quantum
  Circuits}},}\ }\href {http://doi.org/10.1103/PhysRevLett.118.057703}
  {\bibfield  {journal} {\bibinfo  {journal} {Phys. Rev. Lett.}\ }\textbf
  {\bibinfo {volume} {118}},\ \bibinfo {pages} {057703} (\bibinfo {year}
  {2017})}\BibitemShut {NoStop}%
\bibitem [{\citenamefont {Quintana}\ \emph {et~al.}(2017)\citenamefont
  {Quintana}, \citenamefont {Chen}, \citenamefont {Sank}, \citenamefont
  {Petukhov}, \citenamefont {White}, \citenamefont {Kafri}, \citenamefont
  {Chiaro}, \citenamefont {Megrant}, \citenamefont {Barends}, \citenamefont
  {Campbell}, \citenamefont {Chen}, \citenamefont {Dunsworth}, \citenamefont
  {Fowler}, \citenamefont {Graff}, \citenamefont {Jeffrey}, \citenamefont
  {Kelly}, \citenamefont {Lucero}, \citenamefont {Mutus}, \citenamefont
  {Neeley}, \citenamefont {Neill}, \citenamefont {O'Malley}, \citenamefont
  {Roushan}, \citenamefont {Shabani}, \citenamefont {Smelyanskiy},
  \citenamefont {Vainsencher}, \citenamefont {Wenner}, \citenamefont {Neven},\
  and\ \citenamefont {Martinis}}]{Quintana2017}%
  \BibitemOpen
  \bibfield  {author} {\bibinfo {author} {\bibfnamefont {C.~M.}\ \bibnamefont
  {Quintana}}, \bibinfo {author} {\bibfnamefont {Y.}~\bibnamefont {Chen}},
  \bibinfo {author} {\bibfnamefont {D.}~\bibnamefont {Sank}}, \bibinfo {author}
  {\bibfnamefont {A.~G.}\ \bibnamefont {Petukhov}}, \bibinfo {author}
  {\bibfnamefont {T.~C.}\ \bibnamefont {White}}, \bibinfo {author}
  {\bibfnamefont {D.}~\bibnamefont {Kafri}}, \bibinfo {author} {\bibfnamefont
  {B.}~\bibnamefont {Chiaro}}, \bibinfo {author} {\bibfnamefont
  {A.}~\bibnamefont {Megrant}}, \bibinfo {author} {\bibfnamefont
  {R.}~\bibnamefont {Barends}}, \bibinfo {author} {\bibfnamefont
  {B.}~\bibnamefont {Campbell}}, \bibinfo {author} {\bibfnamefont
  {Z.}~\bibnamefont {Chen}}, \bibinfo {author} {\bibfnamefont {A.}~\bibnamefont
  {Dunsworth}}, \bibinfo {author} {\bibfnamefont {A.~G.}\ \bibnamefont
  {Fowler}}, \bibinfo {author} {\bibfnamefont {R.}~\bibnamefont {Graff}},
  \bibinfo {author} {\bibfnamefont {E.}~\bibnamefont {Jeffrey}}, \bibinfo
  {author} {\bibfnamefont {J.}~\bibnamefont {Kelly}}, \bibinfo {author}
  {\bibfnamefont {E.}~\bibnamefont {Lucero}}, \bibinfo {author} {\bibfnamefont
  {J.~Y.}\ \bibnamefont {Mutus}}, \bibinfo {author} {\bibfnamefont
  {M.}~\bibnamefont {Neeley}}, \bibinfo {author} {\bibfnamefont
  {C.}~\bibnamefont {Neill}}, \bibinfo {author} {\bibfnamefont {P.~J.~J.}\
  \bibnamefont {O'Malley}}, \bibinfo {author} {\bibfnamefont {P.}~\bibnamefont
  {Roushan}}, \bibinfo {author} {\bibfnamefont {A.}~\bibnamefont {Shabani}},
  \bibinfo {author} {\bibfnamefont {V.~N.}\ \bibnamefont {Smelyanskiy}},
  \bibinfo {author} {\bibfnamefont {A.}~\bibnamefont {Vainsencher}}, \bibinfo
  {author} {\bibfnamefont {J.}~\bibnamefont {Wenner}}, \bibinfo {author}
  {\bibfnamefont {H.}~\bibnamefont {Neven}}, \ and\ \bibinfo {author}
  {\bibfnamefont {J.~M.}\ \bibnamefont {Martinis}},\ }\bibfield  {title}
  {\enquote {\bibinfo {title} {{Observation of Classical-Quantum Crossover of
  {$1/f$} Flux Noise and Its Paramagnetic Temperature Dependence}},}\ }\href
  {http://doi.org/10.1103/PhysRevLett.118.057702} {\bibfield  {journal}
  {\bibinfo  {journal} {Phys. Rev. Lett.}\ }\textbf {\bibinfo {volume} {118}},\
  \bibinfo {pages} {057702} (\bibinfo {year} {2017})}\BibitemShut {NoStop}%
\bibitem [{\citenamefont {Bylander}\ \emph {et~al.}(2011)\citenamefont
  {Bylander}, \citenamefont {Gustavsson}, \citenamefont {Yan}, \citenamefont
  {Yoshihara}, \citenamefont {Harrabi}, \citenamefont {Fitch}, \citenamefont
  {Cory}, \citenamefont {Nakamura}, \citenamefont {Tsai},\ and\ \citenamefont
  {Oliver}}]{Bylander2011}%
  \BibitemOpen
  \bibfield  {author} {\bibinfo {author} {\bibfnamefont {J.}~\bibnamefont
  {Bylander}}, \bibinfo {author} {\bibfnamefont {S.}~\bibnamefont
  {Gustavsson}}, \bibinfo {author} {\bibfnamefont {F.}~\bibnamefont {Yan}},
  \bibinfo {author} {\bibfnamefont {F.}~\bibnamefont {Yoshihara}}, \bibinfo
  {author} {\bibfnamefont {K.}~\bibnamefont {Harrabi}}, \bibinfo {author}
  {\bibfnamefont {G.}~\bibnamefont {Fitch}}, \bibinfo {author} {\bibfnamefont
  {D.~G.}\ \bibnamefont {Cory}}, \bibinfo {author} {\bibfnamefont
  {Y.}~\bibnamefont {Nakamura}}, \bibinfo {author} {\bibfnamefont {J.-S.}\
  \bibnamefont {Tsai}}, \ and\ \bibinfo {author} {\bibfnamefont {W.~D.}\
  \bibnamefont {Oliver}},\ }\bibfield  {title} {\enquote {\bibinfo {title}
  {{Noise spectroscopy through dynamical decoupling with a superconducting flux
  qubit}},}\ }\href {http://doi.org/10.1038/nphys1994} {\bibfield  {journal}
  {\bibinfo  {journal} {Nat. Phys.}\ }\textbf {\bibinfo {volume} {7}},\
  \bibinfo {pages} {565} (\bibinfo {year} {2011})}\BibitemShut {NoStop}%
\bibitem [{\citenamefont {Anton}\ \emph {et~al.}(2013)\citenamefont {Anton},
  \citenamefont {Birenbaum}, \citenamefont {O'Kelley}, \citenamefont
  {Bolkhovsky}, \citenamefont {Braje}, \citenamefont {Fitch}, \citenamefont
  {Neeley}, \citenamefont {Hilton}, \citenamefont {Cho}, \citenamefont {Irwin},
  \citenamefont {Wellstood}, \citenamefont {Oliver}, \citenamefont {Shnirman},\
  and\ \citenamefont {Clarke}}]{Anton2013}%
  \BibitemOpen
  \bibfield  {author} {\bibinfo {author} {\bibfnamefont {S.~M.}\ \bibnamefont
  {Anton}}, \bibinfo {author} {\bibfnamefont {J.~S.}\ \bibnamefont
  {Birenbaum}}, \bibinfo {author} {\bibfnamefont {S.~R.}\ \bibnamefont
  {O'Kelley}}, \bibinfo {author} {\bibfnamefont {V.}~\bibnamefont
  {Bolkhovsky}}, \bibinfo {author} {\bibfnamefont {D.~A.}\ \bibnamefont
  {Braje}}, \bibinfo {author} {\bibfnamefont {G.}~\bibnamefont {Fitch}},
  \bibinfo {author} {\bibfnamefont {M.}~\bibnamefont {Neeley}}, \bibinfo
  {author} {\bibfnamefont {G.~C.}\ \bibnamefont {Hilton}}, \bibinfo {author}
  {\bibfnamefont {H.-M.}\ \bibnamefont {Cho}}, \bibinfo {author} {\bibfnamefont
  {K.~D.}\ \bibnamefont {Irwin}}, \bibinfo {author} {\bibfnamefont {F.~C.}\
  \bibnamefont {Wellstood}}, \bibinfo {author} {\bibfnamefont {W.~D.}\
  \bibnamefont {Oliver}}, \bibinfo {author} {\bibfnamefont {A.}~\bibnamefont
  {Shnirman}}, \ and\ \bibinfo {author} {\bibfnamefont {J.}~\bibnamefont
  {Clarke}},\ }\bibfield  {title} {\enquote {\bibinfo {title} {{Magnetic Flux
  Noise in dc SQUIDs: Temperature and Geometry Dependence}},}\ }\href {\doibase
  10.1103/PhysRevLett.110.147002} {\bibfield  {journal} {\bibinfo  {journal}
  {Phys. Rev. Lett.}\ }\textbf {\bibinfo {volume} {110}},\ \bibinfo {pages}
  {147002} (\bibinfo {year} {2013})}\BibitemShut {NoStop}%
\bibitem [{\citenamefont {Lanting}\ \emph {et~al.}(2014)\citenamefont
  {Lanting}, \citenamefont {Amin}, \citenamefont {Berkley}, \citenamefont
  {Rich}, \citenamefont {Chen}, \citenamefont {LaForest},\ and\ \citenamefont
  {{de Sousa}}}]{Lanting2014a}%
  \BibitemOpen
  \bibfield  {author} {\bibinfo {author} {\bibfnamefont {T.}~\bibnamefont
  {Lanting}}, \bibinfo {author} {\bibfnamefont {M.~H.}\ \bibnamefont {Amin}},
  \bibinfo {author} {\bibfnamefont {A.~J.}\ \bibnamefont {Berkley}}, \bibinfo
  {author} {\bibfnamefont {C.}~\bibnamefont {Rich}}, \bibinfo {author}
  {\bibfnamefont {S.-F.}\ \bibnamefont {Chen}}, \bibinfo {author}
  {\bibfnamefont {S.}~\bibnamefont {LaForest}}, \ and\ \bibinfo {author}
  {\bibfnamefont {R.}~\bibnamefont {{de Sousa}}},\ }\bibfield  {title}
  {\enquote {\bibinfo {title} {{Evidence for temperature-dependent spin
  diffusion as a mechanism of intrinsic flux noise in SQUIDs}},}\ }\href
  {http://doi.org/10.1103/PhysRevB.89.014503} {\bibfield  {journal} {\bibinfo
  {journal} {Phys. Rev. B}\ }\textbf {\bibinfo {volume} {89}},\ \bibinfo
  {pages} {014503} (\bibinfo {year} {2014})}\BibitemShut {NoStop}%
\bibitem [{\citenamefont {Zaborniak}\ and\ \citenamefont
  {de~Sousa}(2021)}]{Zaborniak2021}%
  \BibitemOpen
  \bibfield  {author} {\bibinfo {author} {\bibfnamefont {T.}~\bibnamefont
  {Zaborniak}}\ and\ \bibinfo {author} {\bibfnamefont {R.}~\bibnamefont
  {de~Sousa}},\ }\bibfield  {title} {\enquote {\bibinfo {title} {{Benchmarking
  Hamiltonian Noise in the D-Wave Quantum Annealer}},}\ }\href {\doibase
  10.1109/tqe.2021.3050449} {\bibfield  {journal} {\bibinfo  {journal} {IEEE
  Trans. Quantum Eng.}\ }\textbf {\bibinfo {volume} {2}},\ \bibinfo {pages}
  {3100206} (\bibinfo {year} {2021})}\BibitemShut {NoStop}%
\bibitem [{\citenamefont {Nava~Aquino}\ and\ \citenamefont
  {de~Sousa}(2022)}]{Nava2022}%
  \BibitemOpen
  \bibfield  {author} {\bibinfo {author} {\bibfnamefont {J.~A.}\ \bibnamefont
  {Nava~Aquino}}\ and\ \bibinfo {author} {\bibfnamefont {R.}~\bibnamefont
  {de~Sousa}},\ }\bibfield  {title} {\enquote {\bibinfo {title} {{Flux noise in
  disordered spin systems }},}\ }\href {\doibase
  https://doi.org/10.1103/PhysRevB.106.144506} {\bibfield  {journal} {\bibinfo
  {journal} {Phys. Rev. B.}\ }\textbf {\bibinfo {volume} {106}},\ \bibinfo
  {pages} {144506} (\bibinfo {year} {2022})}\BibitemShut {NoStop}%
\bibitem [{\citenamefont {Rower}\ \emph {et~al.}()\citenamefont {Rower},
  \citenamefont {Ateshian}, \citenamefont {Li}, \citenamefont {Hays},
  \citenamefont {Bluvstein}, \citenamefont {Ding}, \citenamefont {Kannan},
  \citenamefont {Almanakly}, \citenamefont {Braum\"{u}ller}, \citenamefont
  {Kim}, \citenamefont {Melville}, \citenamefont {Niedzielski}, \citenamefont
  {Schwartz}, \citenamefont {Yoder}, \citenamefont {Orlando}, \citenamefont
  {Wang}, \citenamefont {Gustavsson}, \citenamefont {Grover}, \citenamefont
  {Serniak},\ and\ \citenamefont {Oliver}}]{Rower2023}%
  \BibitemOpen
  \bibfield  {author} {\bibinfo {author} {\bibfnamefont {D.~A.}\ \bibnamefont
  {Rower}}, \bibinfo {author} {\bibfnamefont {L.}~\bibnamefont {Ateshian}},
  \bibinfo {author} {\bibfnamefont {L.~H.}\ \bibnamefont {Li}}, \bibinfo
  {author} {\bibfnamefont {M.}~\bibnamefont {Hays}}, \bibinfo {author}
  {\bibfnamefont {D.}~\bibnamefont {Bluvstein}}, \bibinfo {author}
  {\bibfnamefont {L.}~\bibnamefont {Ding}}, \bibinfo {author} {\bibfnamefont
  {B.}~\bibnamefont {Kannan}}, \bibinfo {author} {\bibfnamefont
  {A.}~\bibnamefont {Almanakly}}, \bibinfo {author} {\bibfnamefont
  {J.}~\bibnamefont {Braum\"{u}ller}}, \bibinfo {author} {\bibfnamefont
  {D.~K.}\ \bibnamefont {Kim}}, \bibinfo {author} {\bibfnamefont
  {A.}~\bibnamefont {Melville}}, \bibinfo {author} {\bibfnamefont {B.~M.}\
  \bibnamefont {Niedzielski}}, \bibinfo {author} {\bibfnamefont {M.~E.}\
  \bibnamefont {Schwartz}}, \bibinfo {author} {\bibfnamefont {J.~L.}\
  \bibnamefont {Yoder}}, \bibinfo {author} {\bibfnamefont {T.~P.}\ \bibnamefont
  {Orlando}}, \bibinfo {author} {\bibfnamefont {J.~I.}\ \bibnamefont {Wang}},
  \bibinfo {author} {\bibfnamefont {S.}~\bibnamefont {Gustavsson}}, \bibinfo
  {author} {\bibfnamefont {J.~A.}\ \bibnamefont {Grover}}, \bibinfo {author}
  {\bibfnamefont {R.}~\bibnamefont {Serniak}, \bibfnamefont {K.~Comin}}, \ and\
  \bibinfo {author} {\bibfnamefont {W.~D.}\ \bibnamefont {Oliver}},\ }\href
  {https://arxiv.org/abs/2301.07804} {\enquote {\bibinfo {title} {{Evolution of
  $1/f$ Flux Noise in Superconducting Qubits with Weak Magnetic Fields}},}\
  }\Eprint {http://arxiv.org/abs/arXiv:2301.07804} {arXiv:2301.07804}
  \BibitemShut {NoStop}%
\bibitem [{\citenamefont {LaForest}\ and\ \citenamefont {{de
  Sousa}}(2015)}]{LaForest2015}%
  \BibitemOpen
  \bibfield  {author} {\bibinfo {author} {\bibfnamefont {S.}~\bibnamefont
  {LaForest}}\ and\ \bibinfo {author} {\bibfnamefont {R.}~\bibnamefont {{de
  Sousa}}},\ }\bibfield  {title} {\enquote {\bibinfo {title} {{Flux-vector
  model of spin noise in superconducting circuits: Electron versus nuclear
  spins and role of phase transition}},}\ }\href
  {http://doi.org/10.1103/PhysRevB.92.054502} {\bibfield  {journal} {\bibinfo
  {journal} {Phys. Rev. B}\ }\textbf {\bibinfo {volume} {92}},\ \bibinfo
  {pages} {054502} (\bibinfo {year} {2015})}\BibitemShut {NoStop}%
\bibitem [{Not()}]{NoteSymmetry}%
  \BibitemOpen
  \href@noop {} {}\bibinfo {note} {This follows from the
  fluctuation-dissipation theorem for spins, Eq.~(7) in \cite{Nava2022}, and
  the observation that the susceptibility $\chi^{ab}_{jj}(\omega)$ is real when
  $a\neq b$.}\BibitemShut {Stop}%
\bibitem [{\citenamefont {Shnirman}\ \emph {et~al.}(2005)\citenamefont
  {Shnirman}, \citenamefont {Sch{\"{o}}n}, \citenamefont {Martin},\ and\
  \citenamefont {Makhlin}}]{Shnirman2005}%
  \BibitemOpen
  \bibfield  {author} {\bibinfo {author} {\bibfnamefont {A.}~\bibnamefont
  {Shnirman}}, \bibinfo {author} {\bibfnamefont {G.}~\bibnamefont
  {Sch{\"{o}}n}}, \bibinfo {author} {\bibfnamefont {I.}~\bibnamefont {Martin}},
  \ and\ \bibinfo {author} {\bibfnamefont {Y.}~\bibnamefont {Makhlin}},\
  }\bibfield  {title} {\enquote {\bibinfo {title} {{Low- and high-frequency
  noise from coherent two-level systems}},}\ }\href {\doibase
  10.1103/PhysRevLett.94.127002} {\bibfield  {journal} {\bibinfo  {journal}
  {Phys. Rev. Lett.}\ }\textbf {\bibinfo {volume} {94}},\ \bibinfo {pages}
  {127002} (\bibinfo {year} {2005})}\BibitemShut {NoStop}%
\bibitem [{\citenamefont {Belli}, \citenamefont {Fanciulli},\ and\
  \citenamefont {de~Sousa}(2020)}]{Belli2020}%
  \BibitemOpen
  \bibfield  {author} {\bibinfo {author} {\bibfnamefont {M.}~\bibnamefont
  {Belli}}, \bibinfo {author} {\bibfnamefont {M.}~\bibnamefont {Fanciulli}}, \
  and\ \bibinfo {author} {\bibfnamefont {R.}~\bibnamefont {de~Sousa}},\
  }\bibfield  {title} {\enquote {\bibinfo {title} {{Probing two-level systems
  with electron spin inversion recovery of defects at the Si/SiO$_2$
  interface}},}\ }\href {\doibase 10.1103/physrevresearch.2.033507} {\bibfield
  {journal} {\bibinfo  {journal} {Phys. Rev. Res.}\ }\textbf {\bibinfo {volume}
  {2}},\ \bibinfo {pages} {033507} (\bibinfo {year} {2020})}\BibitemShut
  {NoStop}%
\bibitem [{\citenamefont {Van~Vleck}(1940)}]{VanVleck1940}%
  \BibitemOpen
  \bibfield  {author} {\bibinfo {author} {\bibfnamefont {J.~H.}\ \bibnamefont
  {Van~Vleck}},\ }\bibfield  {title} {\enquote {\bibinfo {title} {Paramagnetic
  relaxation times for titanium and chrome alum},}\ }\href {\doibase
  10.1103/PhysRev.57.426} {\bibfield  {journal} {\bibinfo  {journal} {Phys.
  Rev.}\ }\textbf {\bibinfo {volume} {57}},\ \bibinfo {pages} {426} (\bibinfo
  {year} {1940})}\BibitemShut {NoStop}%
\bibitem [{\citenamefont {de~Sousa}\ and\ \citenamefont {{Das
  Sarma}}(2003)}]{deSousa2003b}%
  \BibitemOpen
  \bibfield  {author} {\bibinfo {author} {\bibfnamefont {R.}~\bibnamefont
  {de~Sousa}}\ and\ \bibinfo {author} {\bibfnamefont {S.}~\bibnamefont {{Das
  Sarma}}},\ }\bibfield  {title} {\enquote {\bibinfo {title} {{Gate control of
  spin dynamics in III-V semiconductor quantum dots}},}\ }\href {\doibase
  10.1103/PhysRevB.68.155330} {\bibfield  {journal} {\bibinfo  {journal} {Phys.
  Rev. B}\ }\textbf {\bibinfo {volume} {68}},\ \bibinfo {pages} {155330}
  (\bibinfo {year} {2003})}\BibitemShut {NoStop}%
\end{thebibliography}%

\end{document}